%% file: emnlp22.tex
\pdfoutput=1

\documentclass[11pt]{article}

\usepackage{EMNLP2022}

\usepackage{times}
\usepackage{latexsym}

\usepackage[T1]{fontenc}

\usepackage[utf8]{inputenc}

\usepackage{microtype}

\usepackage{inconsolata}

\usepackage{color}
\usepackage{times}
\usepackage{soul}
\usepackage{url}
\usepackage{caption}
\usepackage{graphicx}
\usepackage{amsmath}
\usepackage{amsthm}
\usepackage{booktabs}
\usepackage{algorithm}
\usepackage{algorithmicx}
\usepackage{algpseudocode}
\usepackage{wrapfig}
\usepackage{multirow}
\usepackage{subfig}
\usepackage{float}
\newcommand{\eg}{e.g.,\xspace}
\newcommand{\ie}{i.e.,\xspace}

\usepackage{newfloat}
\usepackage{amsfonts}

\input{math_commands.tex}

\usepackage{times}
\usepackage{url}
\usepackage{etoolbox,multirow,adjustbox}
\usepackage{wrapfig,lipsum,booktabs}
\usepackage{hyperref}       
\usepackage{url}            
\usepackage{booktabs}       
\usepackage{nicefrac}       
\usepackage{microtype}      
\usepackage{multirow}
\usepackage{subfig}
\usepackage{enumerate}
\usepackage{caption}
\usepackage{graphicx}
\usepackage{xspace}
\usepackage{paralist}
\usepackage{float}
\usepackage{scrextend}
\usepackage[tight-spacing=true]{siunitx}
\usepackage{color,soul}
\usepackage{etoolbox,multirow,adjustbox}
\usepackage{comment}
\usepackage{enumitem,kantlipsum}
\usepackage{xcolor}
\usepackage{makecell}

\newlength{\oldintextsep}
\usepackage{amsthm}
\usepackage{algpseudocode,algorithm}
\usepackage{tabularx,adjustbox}
\newcolumntype{L}[1]{>{\hsize=#1\hsize\raggedright\arraybackslash}X}%
\newcolumntype{R}[1]{>{\hsize=#1\hsize\raggedleft\arraybackslash}X}%
\newcolumntype{C}[1]{>{\hsize=#1\hsize\centering\arraybackslash}X}%
\usepackage{todonotes}
\usepackage{breqn}
\clearpage
\extrafloats{100}
\RequirePackage{tcolorbox}
\tcbuselibrary{breakable}
\tcbset{parbox=false, left=1ex, right=1ex}

\def\eg{{\em e.g.,}\xspace}
\def\ie{{\em i.e.,}\xspace}
\def\cf {{\em c.f.,}\xspace}

\newcommand{\tabref}[1]{Table~\ref{#1}\xspace}

\newcommand{\resource}[1]{\textsc{#1}}

\title{Extracted BERT Model Leaks More Information than You Think!}
\author{
    Xuanli He$^1$\Thanks{ Equal contribution. Most of the work was finished when X.H. was at Monash University. Work done during C.C.'s internship at Sony AI.}, Chen Chen$^{2*}$, Lingjuan Lyu$^3$\Thanks{ Corresponding author.}, Qiongkai Xu$^4$ \\
     
    $^1$University College London,
  $^2$Zhejiang University,
  $^3$Sony AI, $^4$ The University of Melbourne \\
  \texttt{h.xuanli@ucl.ac.uk, cc33@zju.edu.cn}\\ \texttt{Lingjuan.Lv@sony.com, Qiongkai.Xu@unimelb.edu.au} \\
}

\begin{document}
\maketitle
\begin{abstract}
The collection and availability of big data, combined with advances in pre-trained models (e.g. BERT), have revolutionized the predictive performance of natural language processing tasks. This allows corporations to provide machine learning as a service (MLaaS) by encapsulating fine-tuned BERT-based models as APIs. Due to significant commercial interest, there has been a surge of attempts to steal remote services via model extraction. Although previous works have made progress in defending against model extraction attacks, there has been little discussion on their performance in preventing privacy leakage. This work bridges this gap by launching an attribute inference attack against the extracted BERT model. Our extensive experiments reveal that model extraction can cause severe privacy leakage even when victim models are facilitated with advanced defensive strategies.
\end{abstract}

\section{Introduction}
The emergence of pre-trained language models (PLMs) has revolutionized the natural language processing (NLP) research, leading to state-of-the-art (SOTA) performance on a wide range of tasks~\cite{devlin2018bert,yang2019xlnet}. This breakthrough has enabled commercial companies to deploy machine learning models as black-box APIs on their cloud platforms to serve millions of users, such as Google Prediction API\footnote{https://cloud.google.com/prediction}, Microsoft Azure Machine Learning\footnote{https://studio.azureml.net}, and Amazon Machine Learning\footnote{https://aws.amazon.com/machine-learning}.

However, recent works have shown that existing NLP APIs are vulnerable to model extraction attack (MEA), which can reconstruct a copy of the remote NLP model based on the carefully-designed queries and outputs of the target API \cite{krishna2019thieves,wallace2020imitation}, causing the financial losses of the target API. 
Prior to our work,  researchers have investigated the hazards of model extraction under various settings, including stealing commercial APIs~\cite{wallace2020imitation, xu2021beyond}, ensemble model extraction~\cite{xu2021beyond}, and adversarial examples transfer~\cite{wallace2020imitation,he2021model}.

Previous works have indicated that an adversary can leverage the extracted model to conduct adversarial example transfer, such that these examples can corrupt the predictions of the victim model~\cite{wallace2020imitation,he2021model}. Given the success of MEA and adversarial example transfer, we conjecture that the predictions from a victim model could reveal its private information unconsciously, as victim models can memorize side information in addition to the task-related message~\cite{lyu2020differentially_SIGIR,lyu2020differentially,carlini2021extracting}. Thus, we are interested in examining whether the victim model can leak the private information of its data to the extracted model, which has received little attention in previous research. In addition, a list of defenses against MEA has been devised~\cite{lee2018defending, ma2021undistillable, xu2021beyond, he2021protecting,he2022cater}. Although these technologies can alleviate the effects of MEA, it is unknown whether such defenses can prevent private information leakage, \eg gender, age, identity.

To study the privacy leakage from MEA, we first leverage MEA to obtain a white-box extracted model. Then, we demonstrate that from the extracted model, it is possible to infer sensitive attributes of the data used by the victim model. To the best of our knowledge, this is the first attempt that investigates privacy leakage from the extracted model. Moreover, 
we demonstrate that the privacy leakage is resilient to advanced defense strategies even though the task utility of the extracted model is significantly diminished, which could motivate further investigation on defense technology in MEA.\footnote{Code and data are available at: \url{https://github.com/xlhex/emnlp2022_aia.git}}


\section{Related Work}
MEA aims to steal an intellectual model from cloud services~\cite{tramer2016stealing,orekondy2019knockoff,krishna2019thieves,wallace2020imitation}.
It has been studied both empirically and theoretically, on simple classification tasks~\cite{tramer2016stealing}, vision tasks~\cite{orekondy2019knockoff},
and NLP tasks~\cite{krishna2019thieves,wallace2020imitation}.
MEA targets at imitating the functionality of a black-box victim model~\cite{krishna2019thieves,orekondy2019knockoff}, \ie a model replicating the performance of the victim model.

Furthermore, the extracted model could be used as a reconnaissance step to facilitate later attacks~\cite{krishna2019thieves}. For instance, the adversary could construct transferrable adversarial examples over the extracted model to corrupt the predictions of the victim model~\cite{wallace2020imitation,he2021model}. Prior works~\cite{coavoux2018privacy,lyu2020differentially} have shown malicious users can infer confidential attributes based on the interaction with a trained model. However, to the best of our knowledge, none of the previous works investigate whether the extracted model can facilitate privacy leakage of the data used by the black-box victim model. 

In conjunction with MEA, a list of avenues has been proposed to defend against MEA. These approaches focus on the perturbation of the posterior prediction. \citet{orekondy2019knockoff} suggested revealing the top-K posterior probabilities only. \citet{lee2018defending} demonstrated that API owners could increase the difficulty of MEA by softening the posterior probabilities and imposing a random noise on the non-argmax probabilities. \citet{ma2021undistillable} introduced an adversarial training process to discourage the knowledge distillation from the victim model to the extracted model. However, these approaches are specific to model extraction, which are not effective to defend against attribute inference attack, as shown in~\secref{sec:defense}.

\section{Attacking BERT-based API}
We first describe the process of MEA. Then we detail the proposed attack: \emph{attribute inference attack} (AIA). Throughout this paper, we mainly focus on the BERT-based API as the victim model, which is widely used in commercial black-box APIs.

\begin{figure}
    \centering
    \includegraphics[width=3in]{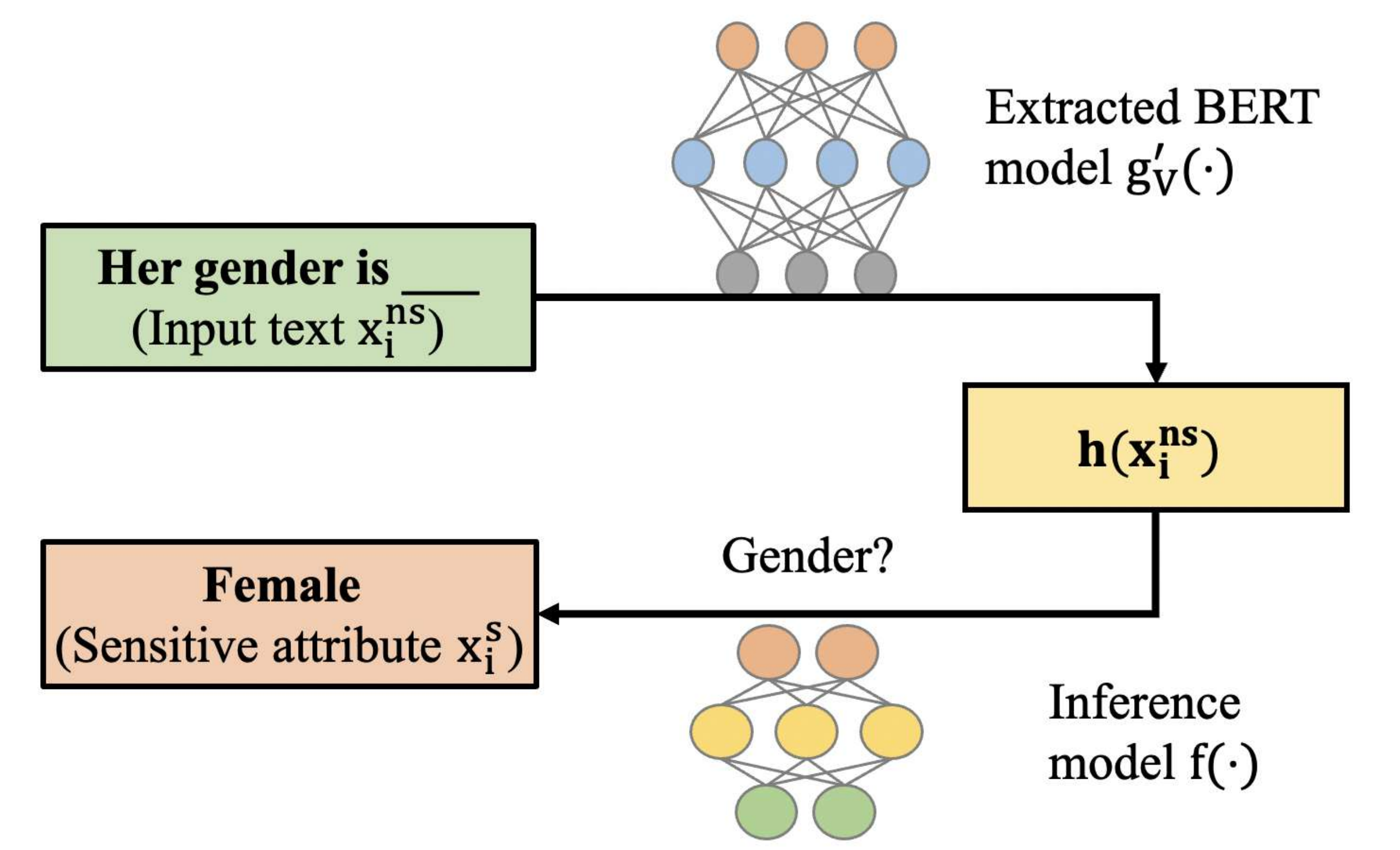}
    \caption{
    The workflow of attribute inference attack against an extracted BERT model. We use an auxiliary attribute inference model to infer the demographic information of a text.} 
    \label{fig:attack_pipeline}
    \vspace{-0.5cm}
\end{figure}

\paragraph{Model Extraction Attack (MEA).} To conduct MEA,  
attackers craft a set of inputs as queries (transfer set), and send them to the target victim model (BERT-based API) to obtain the predicted posterior probability, \ie the outputs of the softmax layer.
Then attackers can reconstruct a copy of the victim model as an ``extracted model'' by training on query-prediction pairs.


\paragraph{Attribute Inference Attack (AIA).}
\label{sec3:AIA}
After we derive an extracted model, we now investigate how to infer sensitive information from the extracted model by 
conducting AIA against the extracted model. 
Given any record $\vx=[x^{ns},x^s]$, AIA aims to reconstruct the sensitive components $x^s$, based on the hidden representation of $x^{ns}$, 
where $x^{ns}$ and $x^s$ represent the non-sensitive information and the target sensitive attribute respectively. The intuition behind AIA is that the representation generated by the extracted model can be used to facilitate the inference of the sensitive information of the data used by the victim model~\cite{coavoux2018privacy}. Note that the \textbf{only} explicit information that is accessible to the attacker is the predictions output by the victim model, rather than the raw BERT representations.



Given an extracted model $g'_V$, we first feed a limited amount of the auxiliary data $D_{aux}$ with labelled attribute into $g'_V$ to collect the BERT representation $\vh(x_i^{ns})$ for each $x_i \in D_{aux}$. 
Then, we train an inference model $f(\cdot)$, which takes the BERT representation of the extracted model as input and outputs the sensitive attribute of the input, i.e., $\{\vh(x_i^{ns}), x_i^{s}\}$.
The trained inference model $f(\cdot)$ can infer the sensitive attribute; in our case, they are gender, age and named entities (see~\secref{sec:data}). 

During test time, as illustrated in Figure~\ref{fig:attack_pipeline}, the attacker can first derive the BERT representation of any record by using the extracted model, then feed the extracted BERT representation into the trained inference model $f(\cdot)$ to infer the sensitive attributes. 


\section{Experiments and Analysis}

\subsection{Experimental Setup}
\label{sec:data}

\input{tab-mea_full_emnlp}

\paragraph{Data.} We conduct experiments on three datasets: \textit{i)} Trustpilot (TP)~\cite{hovy2015user}, \textit{ii)} AG news~\cite{del2005ranking}, and \textit{iii)} Blog posts (Blog)~\cite{schler2006effects}. To study AIA, we reuse the data pre-processed by \citet{coavoux2018privacy}. For TP, \citet{coavoux2018privacy} use the subset from US users, \ie TP-US. The private attributes of TP-US and Blog are \textit{gender} and \textit{age}. The private attributes of AG news are the five most frequent person entities. More details and statistics are provided in Appendix~\ref{app:data}.


\label{sec:data_partition}

\paragraph{Settings.} For each dataset, we randomly split the training data $D$ into two halves $D_V$ and $D_{Q}$, where $|D_V|=|D_{Q}|$. The first half ($D_V$) is used to train the victim model, whereas the second half ($D_{Q}$) is reserved for two folds.
The first fold is to train an extracted model, where the data distribution of the victim's training data ($\mathcal{T}_V$) is the same as that of the query ($\mathcal{T}_A$). The second fold is to train $f(\cdot)$ to infer the private attributes, \ie $D_{aux}$. 

Since API providers tend to use in-house datasets, it is difficult for the attacker to know the target data distribution as prior knowledge. Thus, we sample queries from different distributions but semantically-coherent data as the original data ($\mathcal{T}_A \neq \mathcal{T}_V$). Specifically, we use Amazon reviews dataset~\cite{zhang2015character} (reviews) and CNN/DailyMail dataset~\cite{hermann2015teaching} (news) as cross-domain queries. Empirically, each query incurs a certain expense. Due to the budget limit, attackers cannot issue massive requests. For the cross-domain case, we vary query size from \{0.1x,1x,5x\} size of $D_V$. 

In order to test AIA, we assume $D_V$ is accessible to attackers. We use the non-sensitive attributes of $D_V$ as the test input. If the attacker can successfully infer the sensitive attributes of $D_V$ given its non-sensitive information, then it will cause a privacy leakage of the victim model. 
Following~\citet{coavoux2018privacy}, for demographic variables (\ie gender and age), we take $1-X$ as \textit{empirical privacy}, 
where $X$ is the average prediction accuracy of the attack models on these two variables; for named entities, we take $1-F$ as \textit{empirical privacy}, where $F$ is the F1 score between the ground truths and the prediction by the attackers on the presence of all the named entities. Higher empirical privacy means lower attack performance.

\paragraph{Training Details.}
Victim and extracted models are \textit{BERT-base}~\cite{devlin2018bert}, trained for 5 epochs with the Adam optimizer~\citep{kingma2014adam} using a learning rate of $2\times10^{-5}$. We use the codebase from Transformers library~\citep{wolf-etal-2020-transformers}. Attribute inference models are 2-layer MLP, trained for 3 epochs with the same optimizer and learning rate. All experiments are run with one Nvidia V100 gpu.

\input{tab-aia_ijcai}
\input{tab-defense}

\paragraph{Baselines.} To gauge the private information leakage, we consider a majority
value for each discrete attribute as a baseline. To evaluate 
how the extracted model 
suffers from AIA, we also take the pretrained BERT without (w/o) fine-tuning as a baseline model to extract representation. Note that BERT (w/o fine-tuning) is a plain model that does not contain any information about the training data of the target model. 

\subsection{Experimental Results}
\label{sec:results}
\textbf{MEA results.} We present the performance of MEA for the same domain querying and cross-domain querying in~\tabref{tab:mea_full}. Due to the domain mismatch, the cross-domain querying underperforms the same-domain querying. Although increasing the cross-domain query size can boost the accuracy of the extracted model, it is still inferior to the same-domain competitor with fewer data. In addition, we notice that AG news prefers \textit{news} data, while \resource{tp-us} and Blog favor \textit{reviews} data. Intuitively, one can attribute this preference to the genre similarity, \ie \textit{news} data is close to AG news, while distant from \resource{tp-us} and Blog. To verify this phenomenon, we calculate the uni-gram and 5-gram overlapping between test sets and different queries in Appendix~\ref{app:data}.

Since we do not have access to the training data of the victim model, we will use \textit{news} data as queries for AG news, and \textit{reviews} data as queries for \resource{tp-us} and Blog, unless otherwise mentioned.

\textbf{AIA results}. We show AIA results using the same-domain and cross-domain queries in \tabref{tab:aia}. \tabref{tab:aia} shows that compared to the BERT (w/o fine-tuning) and majority baselines, the attack model built on the BERT representation of the extracted model 
indeed essentially enhances the attribute inference for the victim training data, \ie more than 3.57-4.97x effective for AG news compared with the baselines, even when using cross-domain queries. The majority baseline is merely a random guess, while BERT (w/o fine-tuning) is a plain model that \textbf{did not} contain any information about the victim training data. However, the extracted model is trained on the queries and the returned predictions from the victim model. This implies that the target model predictions inadvertently capture sensitive information about users, such as their gender, age, and other important attributes, apart from the useful information for the main task.

Interestingly, compared with the queries from the same distribution, \tabref{tab:aia} also shows that queries from different distributions make AIA \textbf{easier} (see the best results corresponding to the lower privacy in bold in \tabref{tab:aia}). We provide a detailed study of this phenomenon in Appendix~\ref{app:sharp}.


\section{Defense}
\label{sec:defense}
Although we primarily focus on the privacy vulnerability of BERT-based APIs in this work, we also test four representative defenses: \textit{i)}: \textbf{Softening predictions:}  Using $\tau$ on softmax layer to scale probability vector~\cite{xu2021beyond}. \textit{ii)}: \textbf{Prediction perturbation:} Adding Gaussian noises with a variance of $\sigma$ to the probability vector~\cite{xu2021beyond}. \textit{iii)}. \textbf{Reverse sigmoid:} Softening the posterior probabilities and injecting a random noise on the non-argmax probabilities~\cite{lee2018defending}. \textit{iv)}. \textbf{Nasty teacher:} Using an adversarial loss to discourage the knowledge distillation from the victim model to the extracted model~\cite{ma2021undistillable}. 
We also propose a new defense called \textbf{Most Least}, in which we set the predicted probabilities of the most and least likely categories to $0.5+\epsilon$ and $0.5-\epsilon$, and zero out others. $\epsilon$ could be set as small as possible. For defense experiment, we set $\epsilon$ to $10^{-5}$.

According to~\tabref{tab:defense}, except \resource{mostleast}, none of the defense avenues can well defend against MEA, unless we significantly compromise the utility (or accuracy) of the victim model. However, such degradation is more detrimental to the victim model than the extracted model. Consequently, the extracted model may surpass the victim model.

Regarding AIA, although \resource{mostleast} manages to defend against MEA, it still falls short of preventing privacy leakage from the extracted model (\cf \tabref{tab:aia} and~\ref{tab:defense}). Among these defenses, merely the hard-labeling ($\tau=0.0$) can slightly mitigate the information leakage caused by AIA. In addition, some defenses, such as prediction perturbation and reverse sigmoid, can even exacerbate privacy leakage. Given that these methods have been used to defend against MEA, such a side effect requires more investigation before it causes a severe implication. We leave this for future study.

\section{Conclusions}
This work reveals that the hazards of the extracted model have been underestimated. In addition to the violation of IP, it can vastly exacerbate the privacy leakage even under challenging scenarios (\eg limited query budget; queries from distributions that are different from that of the training data used by the victim APIs). Such a vulnerability cannot be alleviated by the strong defensive strategies 
against model extraction. We hope our work can raise the alarm for more investigations to the vulnerability of model extraction attack.

\section*{Limitations}
Although our work has revealed the vulnerability of model extraction through a lens of privacy leakage, we have not proposed an effective enough defense approach for AIA. Thus, we encourage the community to investigate this direction to mitigate the adverse social impacts caused by this attack.

\section*{Statement of Ethics}
There are two major ethical issues in this work. The first one is the violation of intellectual property, as model extraction attacks can illegally replicate commercial APIs. The second relates to privacy leakage in model extraction attacks. Both can bring severe ethical concerns to the community when deploying machine learning services on the cloud platform. Although we have shown that some defensive avenues can partly mitigate their vulnerabilities, more efforts should be dedicated to them in future work. 



\bibliographystyle{acl_natbib}
\bibliography{sample-base}

\appendix

\input{appendix}

\end{document}

%% file: math_commands.tex
\usepackage{amsmath,amsfonts,bm}

\def\figref#1{Figure~\ref{#1}}

\def\secref#1{Section~\ref{#1}}

\def\eqref#1{equation~\ref{#1}}

\def\1{\bm{1}}

\def\vh{{\bm{h}}}

\def\vx{{\bm{x}}}

\DeclareMathAlphabet{\mathsfit}{\encodingdefault}{\sfdefault}{m}{sl}
\SetMathAlphabet{\mathsfit}{bold}{\encodingdefault}{\sfdefault}{bx}{n}



%% file: tab-mea_full_emnlp.tex
\begin{table}[t]
    \centering
    \scalebox{0.85}{
    \begin{tabular}{c|ccc}
    \hline
       & AG news & Blog & TP-US \\
         \hline
     Victim model & 79.99 & 97.07 &  85.53 \\
         \hline
        \makecell{$\mathcal{T}_A=\mathcal{T}_V$ ($D_{Q}$)} &  \textbf{80.10} & \textbf{95.64} & \textbf{86.53} \\
        \hline

       \makecell{$\mathcal{T}_A \neq \mathcal{T}_V$ (reviews)}  \\
       0.1x & 50.90 & 36.83 & 79.95 \\
       1x & 69.94 & 88.16 &  85.15  \\
       5x & 75.29 & 92.75 & 85.82 \\
       \hline
        \makecell{$\mathcal{T}_A \neq \mathcal{T}_V$ (news)}  \\
        0.1x & 61.70 & 18.04 & 79.20 \\
        1x & 71.95  & 83.13	&  84.15 \\
        5x & 75.82 & 87.64 & 85.46 \\
        \hline
    \end{tabular}}
    \caption{Performance of MEA across different domains and query sizes on the test set, compared to the victim models. The evaluation metric is accuracy. 
    }
    \label{tab:mea_full}
    \vspace{-0.4cm}
\end{table}

%% file: tab-aia_ijcai.tex
\begin{table}
\centering
\scalebox{0.85}{
     
    \begin{tabular}{cccc}
         \toprule
       & AG news & Blog &  TP-US \\
         \midrule
Majority class & 49.94 & 49.57 & 38.15 \\
\midrule
BERT (w/o fine-tuning)&  69.39 &  44.03 & 49.38\\
\midrule
$\mathcal{T}_A = \mathcal{T}_V$ ($D_{aux}$)
&  15.68 & 34.41 & 36.19\\
\midrule
$\mathcal{T}_A \neq \mathcal{T}_V$ (reviews)\\
0.1x & 20.63 & 35.03& \textbf{35.04}\\
       1x  &   17.93 &  34.34  & 35.97 \\
        5x &  18.31 &  34.45&  36.82 \\
\midrule
$\mathcal{T}_A \neq \mathcal{T}_V$ (news)\\
        0.1x & \textbf{13.95} & 35.60& 35.38\\
        1x  &    15.76 & \textbf{33.88}	& 36.92 \\
        5x  &   17.91	&  35.39	&37.68 \\
\bottomrule
    \end{tabular}}
    \caption{
    Empirical privacy of baselines and under AIA attack over different datasets and settings. The extracted model is trained on the queries from different distributions. 
    Note higher value means better empirical privacy.
    }
    \label{tab:aia}
    \vspace{-0.6cm}
\end{table}

%% file: tab-defense.tex
\begin{table*}[h!]
\centering
\scalebox{0.85}{
    
    \begin{tabular}{llccccccccc}
\hline
\multirow{2}{*}{} & &\multicolumn{3}{c}{\resource{ag} news} & \multicolumn{3}{c}{\resource{blog}}& \multicolumn{3}{c}{\resource{tp-us}}\\
\cmidrule(lr){3-5}  \cmidrule(lr){6-8}  \cmidrule(lr){9-11} 
       & & Utility $\uparrow$ & MEA $\downarrow$  &AIA $\uparrow$ & Utility $\uparrow$ & MEA $\downarrow$  &AIA $\uparrow$ &  Utility $\uparrow$ & MEA $\downarrow$  &AIA $\uparrow$\\
\toprule
\multicolumn{2}{c}{No Defense} &79.99 & 71.95 & 15.76 &\textbf{97.07} & 88.16& 34.34 & 85.53& 85.33 & 36.92 \\
\midrule
\multirow{2}{*}{{\rotatebox{90}{\resource{soft.}}}}
& \quad $\tau=0.0$ & 79.99 & 69.11 & \textbf{22.47} &97.07 & 85.57 & \textbf{35.19} & 85.53& 84.60 & 37.62  \\
& \quad $\tau=0.5$ & 79.99 & 72.32 & 20.78 &97.07 & 85.68 &  34.91& 85.53& 85.10 & \textbf{37.69} \\
& \quad $\tau=5$ &79.99 & 72.48 & 11.32 & 97.07 & 86.73 & 33.80&85.53 & 85.33 & 33.18 \\
\midrule
\multirow{2}{*}{{\rotatebox{90}{\resource{pert.}}}}
& \quad $\sigma=0.05$ &\textbf{80.03} & 71.47 & 14.46  & 96.17 & 85.87 & 34.75 & \textbf{85.83} & 85.09 & 37.43\\

& \quad $\sigma=0.2$  & 79.41 & 71.61 & 12.58 &  95.38 & 85.31 & 34.97 & 85.65 & 84.98 & 36.90\\
& \quad $\sigma=0.5$ &65.13 & 69.05 & 11.66 & 62.23 & 81.77 & 33.79 & 63.21 & 83.88 & 35.90\\
\midrule
\multicolumn{2}{c}{Reverse Sigmoid} & 79.99  &71.59	& 12.17 & 97.07  & 85.08 & 33.09 & 85.53 & 85.34 & 32.81 \\
\midrule
\multicolumn{2}{c}{NASTY} & 79.90 & 71.33 & 17.00	& 96.05 & 85.61 &	34.24	&  85.15 & 84.40	& 36.77\\
\midrule
\multicolumn{2}{c}{\resource{MostLeast}} & 79.99 & \textbf{47.98} &  17.86	&  97.07 & \textbf{48.29} & 34.44 & 85.53 & \textbf{39.40} & 37.60\\
\bottomrule
\end{tabular}%
}
\caption{Attack performance under different defenses on AG News, BLOG and TP-US. $\tau$ is temperature parameter on softmax. $\sigma$ is the variance of Gaussian noise. Utility means the accuracy of the victim model after adopting defense. 
For MEA, \textbf{lower} scores indicate better defenses.  
conversely for AIA. 
All experiments are conducted on datasets with 1x queries.}
    \label{tab:defense}
    \vspace{-0.5cm}
\end{table*}

%% file: appendix.tex
\section{Datasets}
\label{app:data}
This section first details the construction of each dataset.


\paragraph{Trustpilot (TP)}. Trustpilot sentiment dataset~\cite{hovy2015user} contains reviews associated with a sentiment score on a five point scale, and each review is associated with 3 attributes: gender, age and location, which are self-reported by users. The original dataset is comprised of reviews from different locations, however, in this paper, we only derive \resource{tp-us} for study. Following \cite{coavoux2018privacy}, we extract examples containing information of both gender and age, and treat them as the private information.

\input{tab-data_emnlp}

\paragraph{AG news}. \resource{ag} news corpus~\cite{del2005ranking} aims to predict the topic label of the document, with four different topics in total. Following \cite{zhang2015character,jin2019bert}, we use both ``title'' and ``description'' fields as the input document.
We use full \resource{ag} news dataset for MEA, which we call \resource{ag} news (full). As AIA 
takes the entity as the sensitive information, we use the corpus filtered by~\cite{coavoux2018privacy}, which we call \resource{ag} news. The resultant \resource{ag} news merely includes sentences with the five most frequent person entities, and each sentence contains at least one of these named entities. Thus, the attacker can identify these five entities as five independent binary classification tasks.

\input{tab-overlap}
\input{tab-att_acc_emnlp}

\begin{figure*}[t!]
	\centering
	\subfloat[\resource{ag} news]{\includegraphics[width=1.8in]{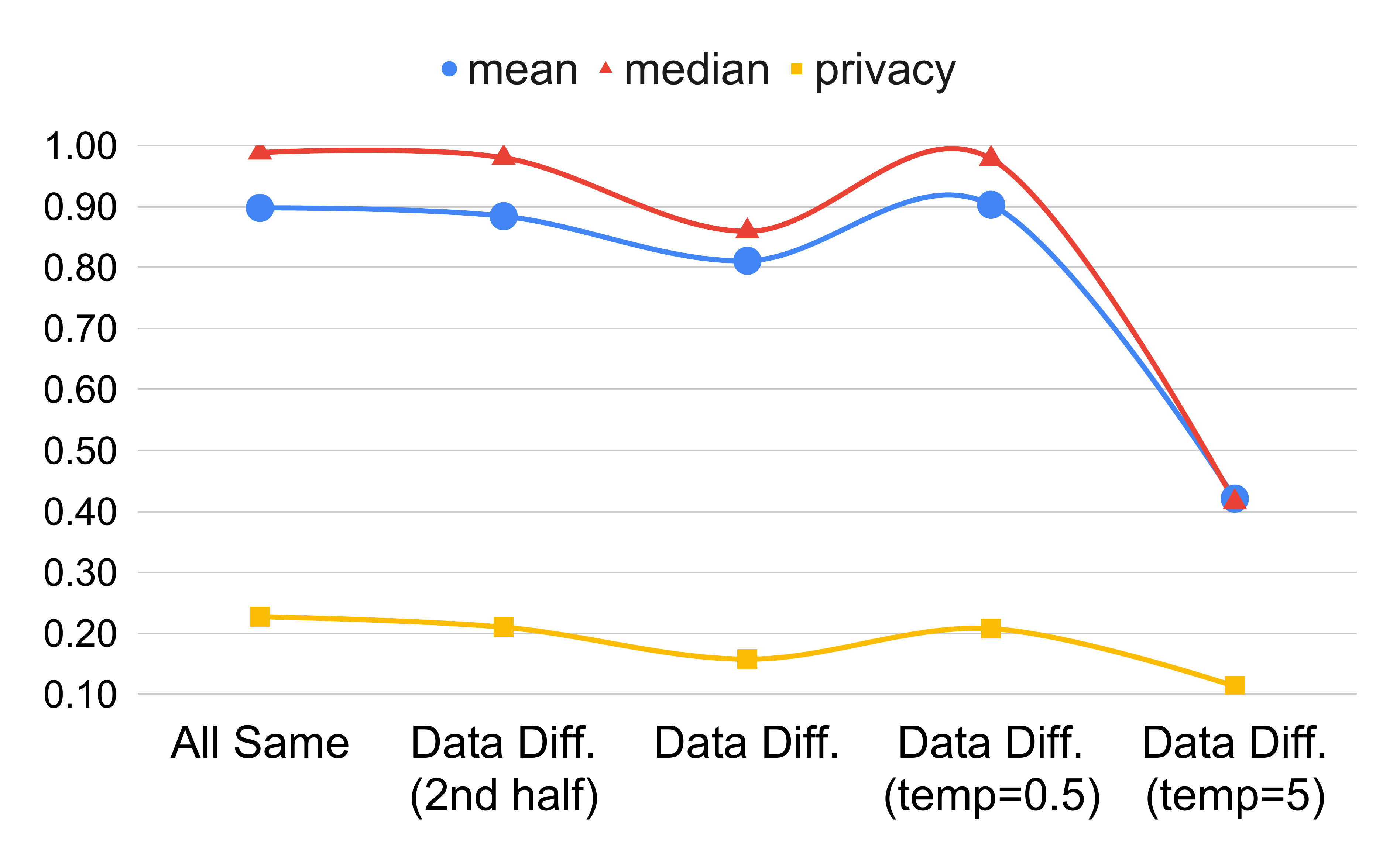}}\ \ \
	\subfloat[\resource{tp-us}]{\includegraphics[width=1.8in]{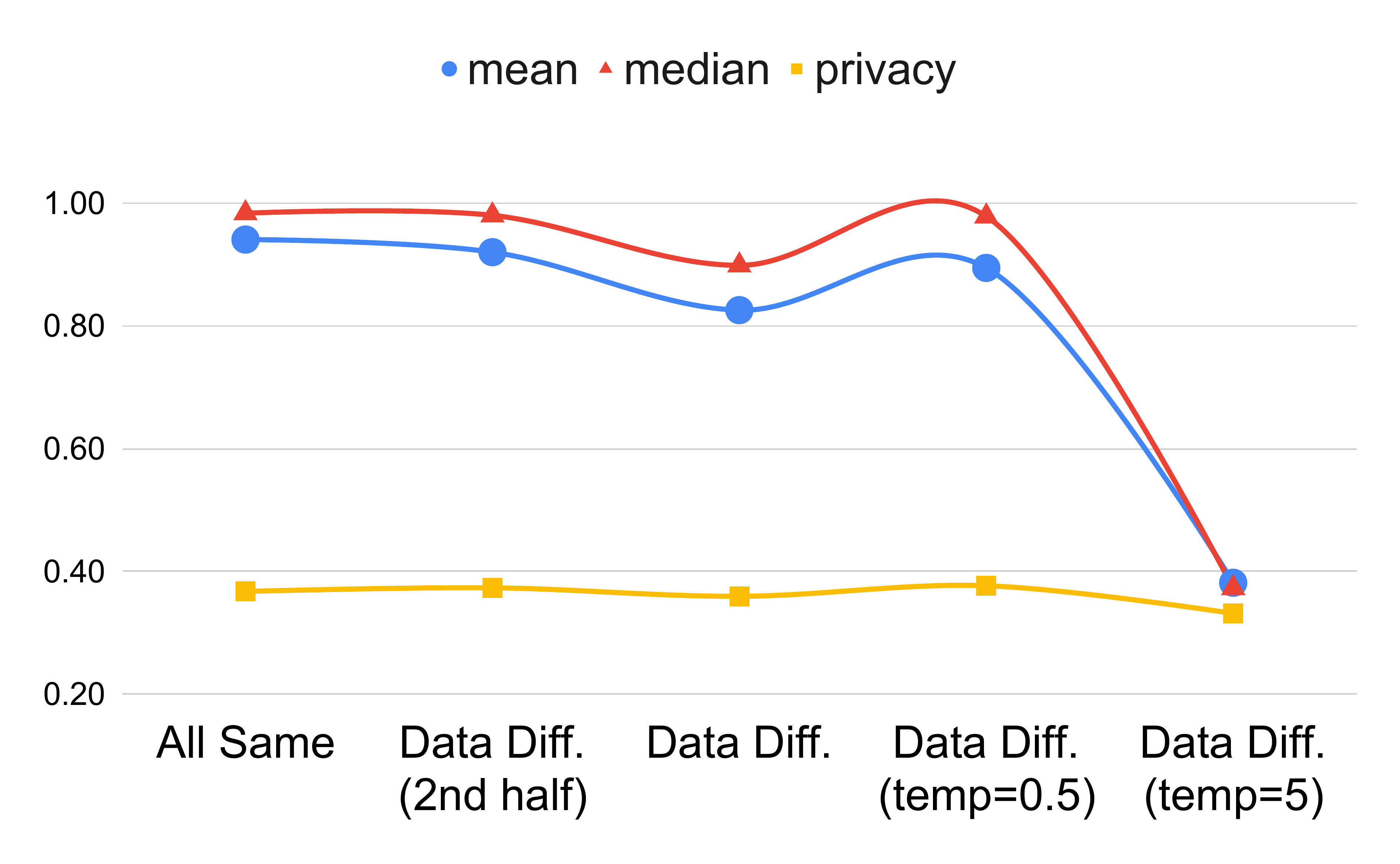}}\ \ \
	\subfloat[Blog]{\includegraphics[width=1.8in]{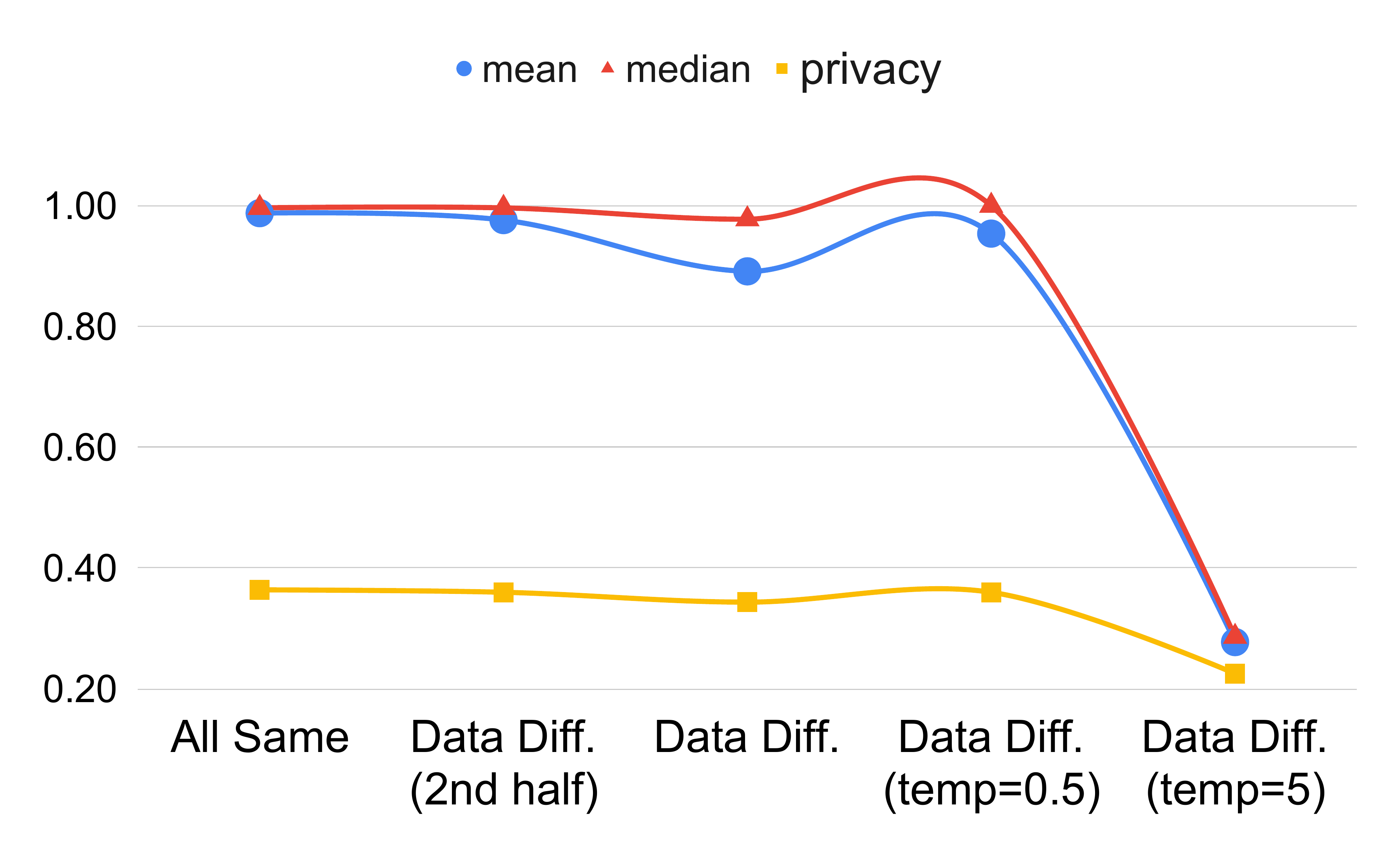}}\\
	\caption{The correlation between the empirical privacy of AIA and the maximum posterior probability. 
	\textbf{mean} and \textbf{median} denote the mean and median of the maximum posterior probability of the queries.}
    \label{fig:aia_maxprob}
\end{figure*}

\paragraph{Blog posts (Blog)}. We derive a blog posts dataset from the blog authorship corpus~\cite{schler2006effects}. We recycle the corpus preprocessed by ~\cite{coavoux2018privacy}, which covers 10 different topics. Similar to \resource{TP-US}, the private variables are comprised of the age and gender of the author.

We provide the statistics of all datasets in~\tabref{tab:data}. \tabref{tab:overlap} presents the uni-gram and 5-gram overlapping between test sets and different queries. According to~\tabref{tab:overlap}, AG news is closer to news data, whereas Blog and TP-US are more similar to reviews data, which validates our claim in~\secref{sec:results}.

\section{Supplementary Studies}
\subsection{Impact of Prediction Sharpness}
\label{app:sharp}
Interestingly, compared with the queries from the same distribution, \tabref{tab:aia} also shows that queries from different distributions make AIA \textbf{easier} (see the best results corresponding to the lower privacy in bold in \tabref{tab:aia}).
We hypothesize this counter-intuitive phenomenon is due to that the posterior probability of the same distribution is sharper than that of the different distribution.
This argument can be further strengthened in Section \ref{sec:defense}, in which we use a temperature coefficient $\tau$ at the softmax layer to control the sharpness of the posterior probability. We conjecture that if the model is \emph{less confident} on its most likely prediction, then AIA is more likely to be successful. This speculation is confirmed by \figref{fig:aia_maxprob}, where the higher posterior probability leads to a higher empirical privacy.

\subsection{Impact of Attribute Distribution}

We further investigate which attribute is more vulnerable, \ie the relationship between attribute distribution (histogram variance) and privacy leakage. 
\tabref{tab:att_acc} empirically indicates that attributes with higher variances cause more information leakage or a lower empirical privacy. For example, for AG-news, entity 2-4 with higher variances result in lower empirical privacy, while entity 0-1 are more resistant to AIA. For TP-US and Blog, as age and gender exhibit similar distribution, AIA performance gap across these two attributes is less obvious.

\subsection{Architecture Mismatch}
In practice, it is more likely that the adversary does not know the victim's model architecture. A natural question is whether our attacks are still possible when the extracted models and the victim models have different architectures, \ie architectural mismatch. To study the influence of the architectural mismatch, we fix the architecture of the extracted model, while varying the victim model from BERT~\cite{devlin2018bert}, RoBERTa~\cite{liu2019roberta} to XLNET~\cite{yang2019xlnet}. According to \tabref{tab:bert_mismatch}, when there is an architecture mismatch between the victim model and the extracted model, the efficacy of AIA is alleviated as expected. However, the leakage of the private information is still severe (compare to  the majority class in \tabref{tab:aia}). Surprisingly, we observe that MEA cannot benefit from a more accurate victim, which is different from the findings in~\cite{hinton2015distilling}. For example, the victim model performs best using XLNET-large, while MEA shows best performance when the victim model uses XLNET-base. We conjecture such difference is ascribed to the distribution mismatch between the training data of the victim model and the queries. We will conduct an in-depth study on this in the future.

\input{tab-arch}

%% file: tab-data_emnlp.tex
\begin{table}[b!]
 \centering
 \scalebox{0.9}{
    \begin{tabular}{lllll}
    \hline
Dataset & Private Variable & \#Train & \#Dev& \#Test  \\
    \toprule
\resource{tp-us} & age, gender & 22,142  & 2,767 & 2,767  \\
\resource{ag} & named entity & 11,657 & 1,457 &1,457  \\
Blog & age, gender & 7,098 &887 &887 \\
      
     \bottomrule
 \end{tabular}
  }
    \caption{Summary of the studied NLP datasets.}
    \label{tab:data}
\end{table}

%% file: tab-overlap.tex
\begin{table*}[h]

    \centering
    \begin{tabular}{lcc|cc|cc}
    \hline
Query   &   \multicolumn{2}{c}{AG news} & \multicolumn{2}{c}{Blog} & \multicolumn{2}{c}{TP-US}\\
         \toprule
         
    &     uni-gram & 5-gram &uni-gram & 5-gram & uni-gram & 5-gram  \\
           \midrule
reviews   & 68.22\% &0.53\% & 47.21\% & 0.73\% & 60.86\% & 2.57\%\  \\
       
news  & 72.13\% & 1.24\% & 44.76\% & 0.06\% & 51.28\% & 0.12\%  \\
    \hline
    \end{tabular}
    \caption{Percentage of uni-gram and 5-gram recall-based overlap between different queries and test sets.
    }
    \label{tab:overlap}
\end{table*}

%% file: tab-att_acc_emnlp.tex
\begin{table*}
\centering
\scalebox{0.8}{
    \begin{tabular}{cccccc}
    \hline
     \multicolumn{6}{c}{\resource{ag} news}\\
     \hline
        	& {\makecell{entity 0 \\(std=310.0)}}	& {\makecell{entity 1 \\(std=1568.5)}}	&{\makecell{entity 2 \\(std=2095.5)}}	& {\makecell{entity 3 \\(std=2640.5)}}	& {\makecell{entity 4 \\(std=2615.5)}}\\
        	\hline
$\mathcal{T}_A=\mathcal{T}_V$	& 15.61 &    15.10	&  7.71	&   6.95	& 5.49 \\
\makecell{$\mathcal{T}_A \neq \mathcal{T}_V$ \\(news)} 
& 14.79	& 12.38	& 3.84	& 5.33	&2.02\\ \hline
& \multicolumn{2}{c|}{\resource{tp-us}} & \multicolumn{2}{c}{Blog} \\
\hline
	& {\makecell{gender \\(std=1512.0)}}	&\multicolumn{1}{c|}{{\makecell{age \\(std=1440.0)}}} 
	&{\makecell{age \\(std=28.0)}}	&{\makecell{gender \\(std=6.0)}} \\
$\mathcal{T}_A=\mathcal{T}_V$	& 36.40& \multicolumn{1}{c|}{37.12}	& 
32.18	& 39.02\\
\makecell{$\mathcal{T}_A \neq \mathcal{T}_V$ \\(reviews)}&	36.44	&\multicolumn{1}{c|}{37.40}	& 
31.20	&38.01\\
\hline
    \end{tabular}}
    \caption{
    AIA performance on attributes of different datasets. All experiments are based on 1x queries. std is the standard deviation of attribute distribution.}
    \label{tab:att_acc}
\end{table*}

%% file: tab-arch.tex

\begin{table}
\begin{center}
\scalebox{0.75}{
\begin{tabular}{lllll} 
 \toprule
 \textbf{Victim} & \textbf{Extracted} & \multicolumn{3}{c}{\textbf{TP-US}}\\
 \cmidrule(lr){3-5}
 & & victim $\uparrow$ &MEA $\uparrow$ & AIA $\downarrow$ \\
 \midrule
 BERT-large & BERT-base & 86.82 & 85.36 &36.65 \\
 RoBERTa-large & BERT-base & 87.20 & 85.72 & 37.33 \\
 RoBERTa-base & BERT-base & 86.66 & 85.40 &37.52 \\
 XLNET-large & BERT-base & 87.21 & 85.99 &37.68 \\
 XLNET-base & BERT-base & 86.91 & \textbf{86.13} &38.09 \\
 \midrule
   BERT-base & BERT-base & 85.53 & 85.15 & \textbf{35.97} \\
 \bottomrule
\end{tabular}}
\end{center}
\caption{
Attack performance on TP-US with mismatched architectures between the victim model and the extracted model. 
}
\label{tab:bert_mismatch}
\vspace{-0.5cm}
\end{table}